\documentclass[a4paper,twocolumn]{article}

\usepackage{amsmath,amssymb,amsfonts}
\usepackage{graphicx}
\usepackage{geometry}
\usepackage{subcaption}
\usepackage{url}
\usepackage{authblk}

\makeatletter

\begin{document}

\onecolumn
\thispagestyle{empty}
This manuscript is a preprint version submitted to IEEE as a conference record of IEEE NSS/MIC 2019.
\par Copylight IEEE 2019.
Personal use of this material is permitted.  
Permission from IEEE must be obtained for all other uses, in any current or future media, including reprinting/republishing this material for advertising or promotional purposes, creating new collective works, for resale or redistribution to servers or lists, or reuse of any copyrighted component of this work in other works.
\clearpage
\addtocounter{page}{-1}
\twocolumn

\title{Front-end-Electronics for the SiPM-readout gaseous TPC for neutrinoless double beta decay search}

\author[1]{K.~Z.~Nakamura\thanks{Corresponding author \\(e-mail: nakamura.kazuhiro.74x@st.kyoto-u.ac.jp)} }
\author[1]{S.~Ban}
\author[1]{A.~K.~Ichikawa}
\author[2]{M.~Ikeno} 
\author[3]{K.~D.~Nakamura} 
\author[1]{T.~Nakaya}
\author[1]{S.~Obara}
\author[1]{S.~Tanaka}
\author[2]{T.~Uchida}
\author[1]{M.~Yoshida}
\affil[1]{Department of Physics, Kyoto University, Kyoto 606-8502, Japan}
\affil[2]{Institute of Particle and Nuclear Studies, High Energy Accelerator Research Organization (KEK), Tsukuba 305-0801, Japan}
\affil[3]{Department of Physics, Kobe University, Kobe 657-8501, Japan}
\date{December 13, 2019}

\maketitle

\begin{abstract}
We have developed a dedicated front-end-electronics board for a high-pressure xenon gas time projection chamber for a neutrinoless double-beta decay search.
The ionization signal is readout by detecting electroluminescence photons with SiPM's.  
The board readout the signal from 56~SiPM's through the DC-coupling and record the waveforms at 5~MS/s with a wide dynamic range up to 7,000~photons/200~ns.  
The SiPM bias voltages are provided by the board and can be adjusted for each SiPM.
In order to calibrate and monitor the SiPM gain, additional auxiliary ADC measures 1~photon-equivalent dark current.
The obtained performance satisfies the requirement for a neutrinoless double-beta decay search.
\end{abstract}

\section{Introduction}
Neutrinoless double beta decay ($0\nu\beta\beta$) is a phenomenon in which two beta-decays simultaneously happen inside a nucleus without emitting neutrinos. 
It happens only in the case neutrino has Majorana-type mass for some nuclei.  
Hence, the observation of $0\nu\beta\beta$ would be a direct proof of the Majorana nature of neutrinos,
but so far it has not been observed and only lower limits on the life have been obtained.
The KamLAND-Zen experiment reported the lower limit on the half life of $0\nu\beta\beta$ as $T_{1/2}^{0\nu} > 1.07\times10^{26}~\mathrm{yr}$ at 90\% C.L.~\cite{PhysRevLett.117.082503}.
The most relevant feature of $0\nu\beta\beta$ is monochromaticity of the energy sum of the two emitted $\beta$-rays. 
The track topology has also characteristic feature: two beta-rays are emitted from a nucleus.
Therefore, in order to conduct the sensitive search for $0\nu\beta\beta$ with good signal and background separation, detectors having high energy resolution and track topology identification has been desired.

\begin{figure}[!t]
    \centering
    \includegraphics[clip,width=\linewidth]{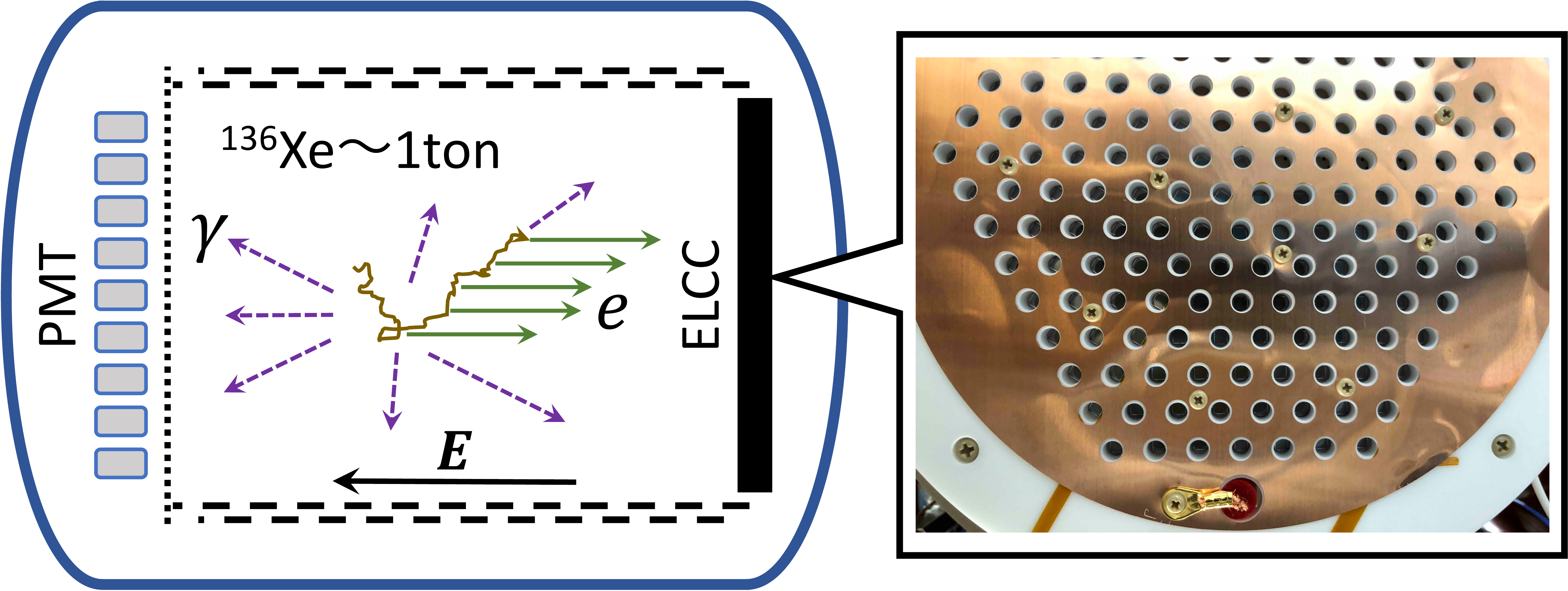}
    \caption{Schematic view of the AXEL detector (left) and ELCC plane (right)}
    \label{fig:AXEL_desctiption}
\end{figure}

We are developing a high-pressure $^{136}\mathrm{Xe}$ gaseous TPC (time projection chamber) detector, AXEL.
Thanks to the pixelized readout utilizing the electroluminescence process, called as ELCC (Electroluminescence Light Collection Cell)~\cite{BAN2017185}, the AXEL detector has good energy resolution and tracking capability. 
The schematic view of AXEL is shown in Fig.~\ref{fig:AXEL_desctiption}. 
Ionization electrons generated by $\beta$-rays are drifted to ELCC and generate electroluminescence (EL) photons by a few $\mathrm{kV/cm/bar}$ electric field applied in each cell.
As the EL process is a linear process~\cite{Aprile:2008bga}, ionization electrons can be converted to the photons with small fluctuation.
EL photons are detected by a SiPM (Hamamatsu MPPC, SS13370) in each cell.
Event track topologies can be reconstructed from the hit pattern and the hit timing of ELCC.
The target energy resolution is $\Delta E = 0.5\mathrm{\%}$ (FWHM) at the Q-value ($2,458 \mathrm{~keV}$) of $^{136}\mathrm{Xe}$.

Currently, we are constructing a 180-L size prototype detector to demonstrate the performance at Q-value and to establish the technique to construct larger-size detectors.
The number of channels of this detector is 1,512.
To operate and readout the waveform of more than 1,500 SiPM's, we have developed a dedicated front-end electronics board, AxFEB (AXEL front-end electronics board).
In this paper, we present the design and performance of the AxFEB.

\section{Design of AxFEB}

\subsection{Requirements to the AxFEB}
\begin{figure}[!t]
    \centering
    \includegraphics[width=\linewidth]{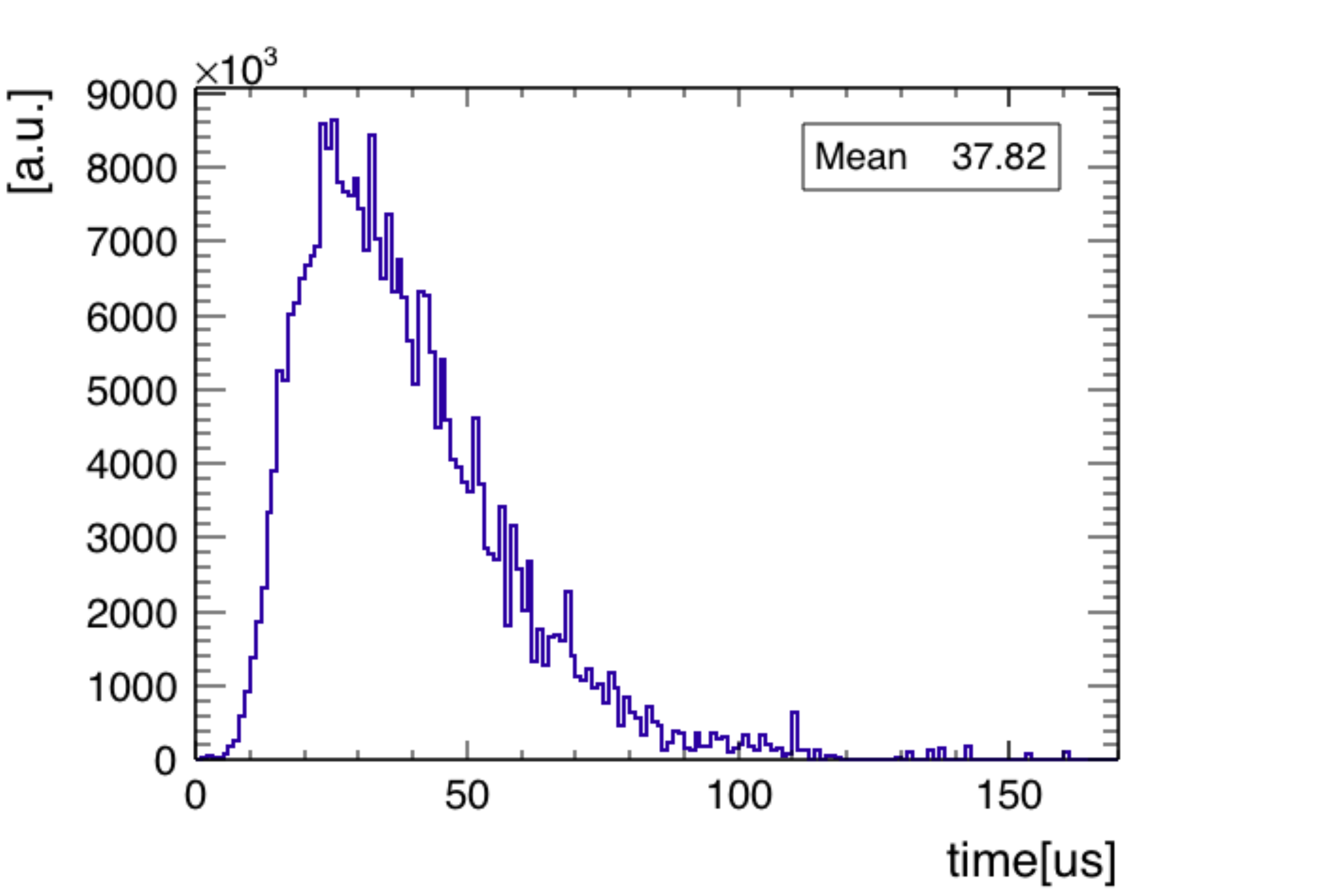}
    \caption{Simulated distribution of signal time width for each channel. For evaluating the effect on the event reconstruction, entries were weighted by the number of signal photons to make this histogram.}
    \label{fig:duration_time}
\end{figure}

\begin{figure}[!t]
    \centering
    \includegraphics[width=\linewidth]{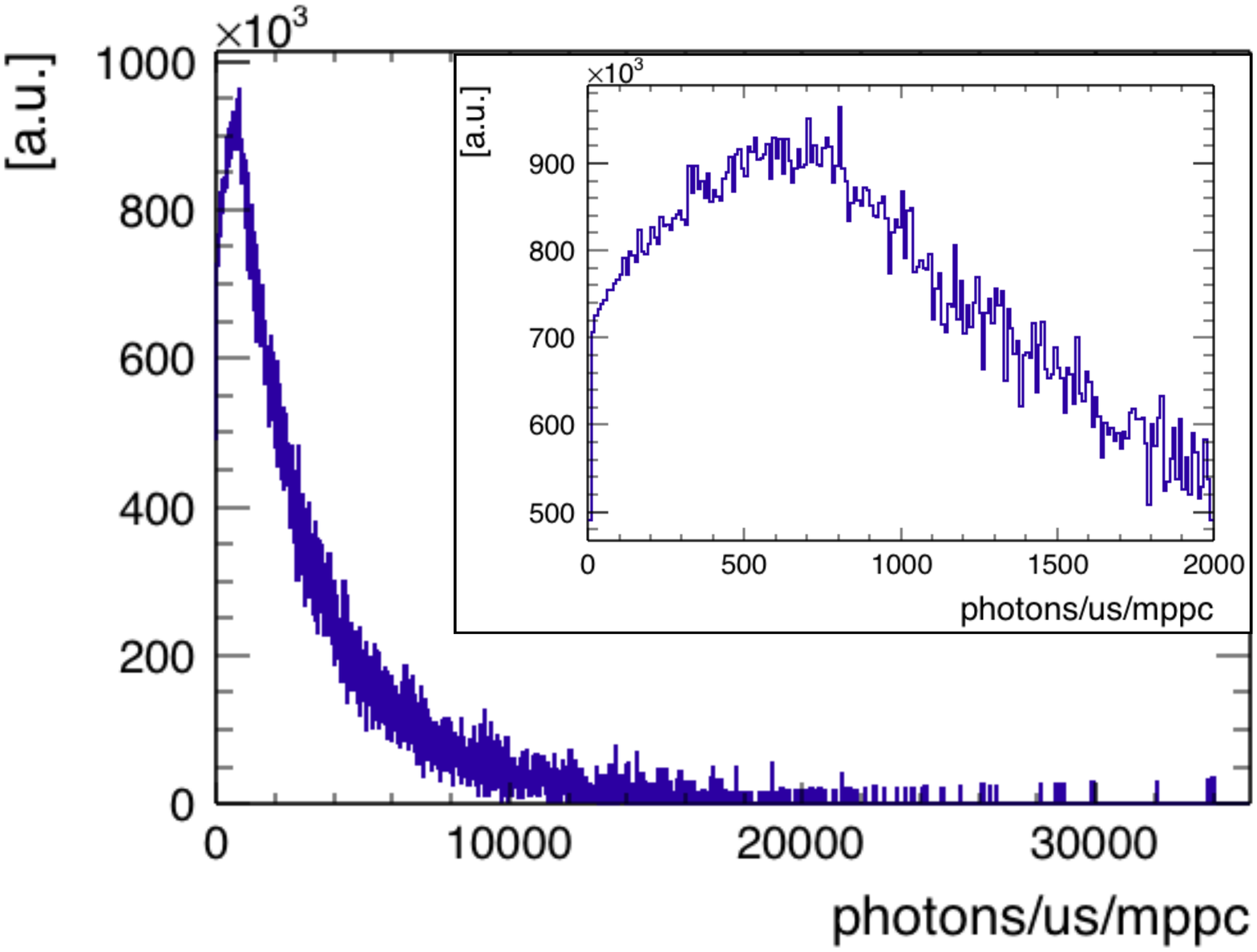}
    \caption{Simulated distribution of photon rate, i.e., number of photons detected by the MPPC in each $1~\mathrm{\mu s}$. Entries were weighted by the number of photons of the signal to make this histogram.}
    \label{fig:photonnum_per1us}
\end{figure}

To evaluate the requirement, $0\nu\beta\beta$ events were simulated by Geant4~\cite{AGOSTINELLI2003250} and Garfield++~\cite{garfieldpp} with responses of ionization electron drift, EL process, and MPPC.
Fig.~\ref{fig:duration_time} shows a distribution of signal time width for each channel obtained by the simulation. 
The required record length is $150~\mathrm{\mu s}$.
The sampling rate required to reconstruct the energy with sufficient resolution was estimated to be $2~\mathrm{MS/s}$. 
However, it turned out that $5~\mathrm{MS/s}$ is necessary to correct the non-linearity of the MPPC. 
Typical electron drift velocity in xenon gas is $\sim 0.1~\mathrm{cm/\mu s}$. 
So $5~\mathrm{MHz}$ corresponds to $0.2~\mathrm{mm}$ in z position sampling, which is smaller than the longitudinal diffusion.
Fig.~\ref{fig:photonnum_per1us} shows distribution of photon rate for each channel, that is, number of photons detected by the MPPC in each $1~\mathrm{\mu s}$.
The photon rate reaches $35,000 ~\mathrm{photons/\mu s}$, but most of the time, the photon rate is less than $2,000~\mathrm{photons/\mu s}$.
With detailed simulation, it was found that required dynamic range is $4$ -- $7,000~\mathrm{photons}/200~\mathrm{ns}$ at $5~\mathrm{MHz}$ sampling to obtain the target energy resolution.

The gain of SiPM is given as following, 
\begin{equation}
    \label{eq:mppc_gain}
    g\equiv \frac{Q_\mathrm{1~p.e.}}{q_{e^-}} = \frac{C(V_\mathrm{bias}-V_\mathrm{break})}{q_{e^-}} = \frac{C\cdot V_\mathrm{over}}{q_{e^-}},
\end{equation}
where, $Q_\mathrm{1~p.e.}$ is the charge of the 1~photon-equivalent ($\mathrm{p.e.}$) signal, $q_{e^-}$ elementary charge, $C$ capacitance of MPPC, $V_\mathrm{bias}$ applied voltage to the MPPC, and $V_\mathrm{break}$ voltage at which APDs (Avalanche Photo Diode) in an MPPC starts causing Geiger discharge.
The $V_\mathrm{over}$ is called overvoltage, and the MPPC gain is proportional to it.
The overvoltage is very small compared to the breakdown voltage, e.g., a typical breakdown voltage of SS13370 is $\simeq 55$ -- $60~\mathrm{V}$ while $V_\mathrm{over} = 3$ -- $4~\mathrm{V}$. 
Besides, the breakdown voltage varies for each MPPC.
Therefore, the front-end board is required to control the bias voltage precisely for each MPPC to get uniform gain for all MPPC's.

The gain calibration and stability monitoring are done by measuring $1~\mathrm{p.e.}$ signal by dark current.
Since the time constant of the $1~\mathrm{p.e.}$ signal is as short as several tens of nanoseconds, it cannot be measured with the $5~\mathrm{MS/s}$ ADC.
Therefore another type of ADC with $40 ~\mathrm{MS/s}$ sampling is added for MPPC calibration.
The typical gain of MPPC S13370 is $2.6\times10^6$, so $1~\mathrm{p.e.}$ signal corresponds to $0.96~\mathrm{pC}$.
To match the $40~\mathrm{MS/S}$, $2~\mathrm{V_{pp}}$, $12~\mathrm{bit}$ ADC whose 1~count corresponds to $0.2~\mathrm{pC}$, the signal is amplified by 165 times in the analog section before being sent to the ADC for calibration.

In summary, the requirements for the front-end electronics circuit are as follows,
\begin{itemize}
    \item Dynamic range: $4$ -- $7,000~\mathrm{photons}/200~\mathrm{ns}$
    \item Record length: $150~\mathrm{\mu s}$
    \item One~$\mathrm{p.e.}$ signal measurement for calibration
    \item Independent bias voltage supply to each MPPC with tens of $\mathrm{mV}$ resolution
    \item Low cost and good extendability for thousand channels.
\end{itemize}

\subsection{Overview of the AxFEB}
\begin{figure}[!t]
    \centering
    \includegraphics[width=\linewidth]{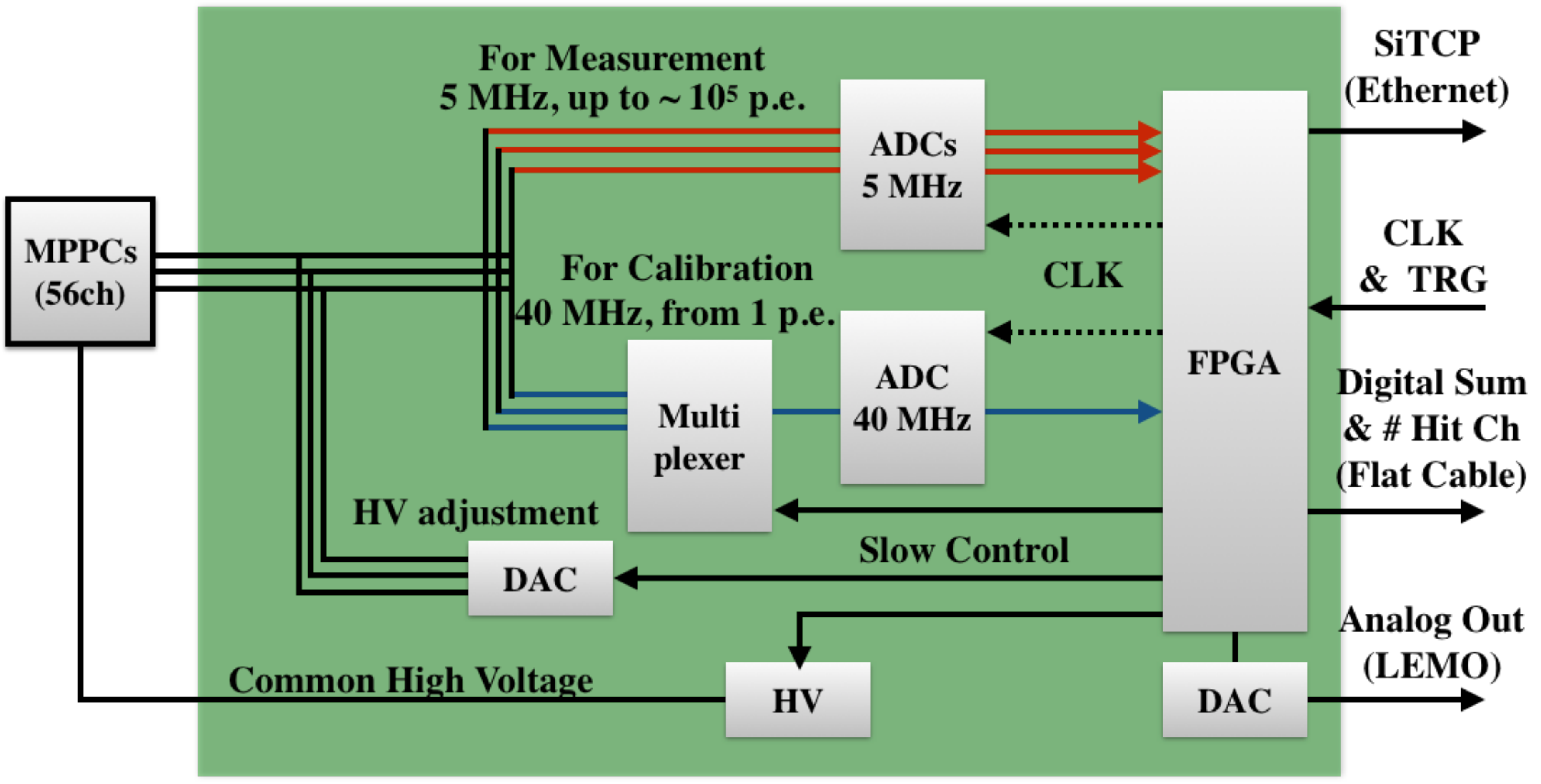}
    \caption{Block diagram of AxFEB}
    \label{fig:block_diagram}
\end{figure}

Fig.~\ref{fig:block_diagram} depicts the block diagram of AxFEB.
One board readout the waveform signal from 56 MPPC's.
The board operates at DC $5\mathrm{~V}$.

Fine adjustments of bias voltage for each MPPC is realized by using two types of DAC's: one for common high voltage supply and is connected to the cathode side of MPPC, and the other for fine adjustment for each MPPC and is connected to the anode side of MPPC.
Special configuration is adopted for analog signal processing to avoid the AC-coupling readout and is described in Sec.~\ref{sec:analog_section}.

There are two types of ADCs in AxFEB -- ADCL and ADCH.
ADCL is used for $0\nu\beta\beta$ detection, and ADCH is used for calibration of MPPC by measuring dark current.
Since calibration only needs to be done intermittently, one ADCL reads out every 7~channels switched by a multiplexer.

The board calculate the sum of 56~channels ADC values and send it to a general trigger module.
The trigger module issues a trigger to each AxFEB according to the sum signal from AxFEBs.
Each AxFEB has an IP address and can communicate via the Ethernet with PC's. 
The overall control, data processing and communication with other modules and PC's are managed by a FPGA on the board.

\subsection{Analog section}
\label{sec:analog_section}
\begin{figure*}[p]
    \centering
   \includegraphics[width=0.9\textheight, angle=270]{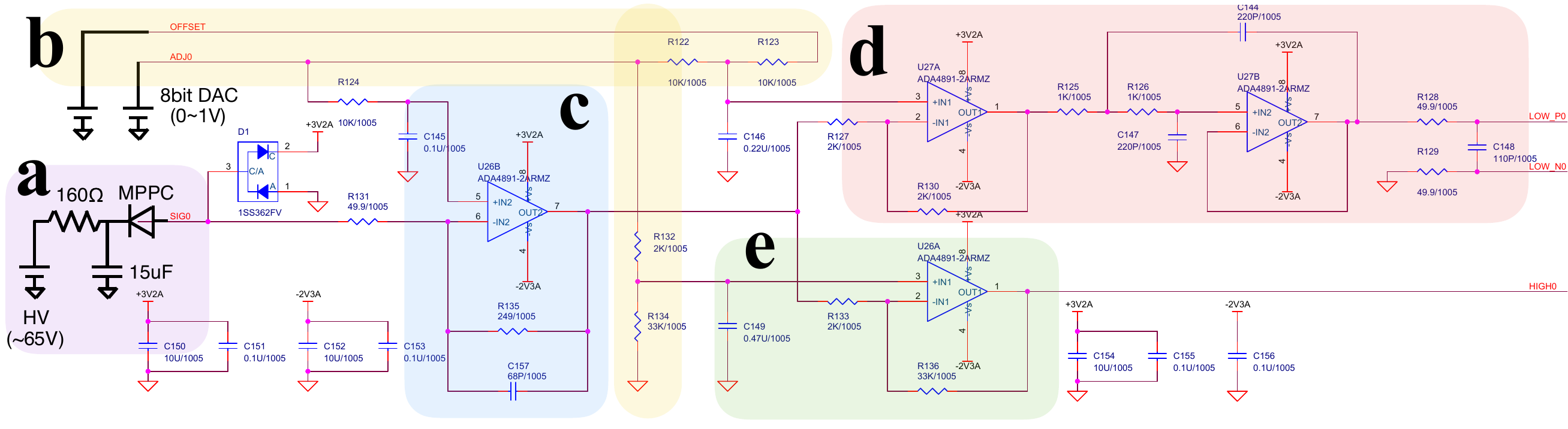}
    \caption{Circuit diagram of the analog part of one channel of the AXELBOAED}
    \label{fig:analog_section}
\end{figure*}

Fig.~\ref{fig:analog_section} shows the diagram of the analog section of AxFEB.
The analog section has three main roles:
\begin{enumerate}
    \item Amplify the signal to match the dynamic range of the ADC
    \item Shaping the signal to match the ADC sampling rate
    \item Apply the bias voltage to each MPPC.
\end{enumerate}

The first-order filter (integration circuit) indicated in Fig.~\ref{fig:analog_section}c shapes the waveform with the $17\mathrm{~ns}$ time constant and also amplifies the signal by 5~times.
The second-order filter (Sallenkey filter) in Fig.~\ref{fig:analog_section}d shapes the waveform to the $\sim 400~\mathrm{ns}$ time constant for ADCL.
The total gain of the circuit for ADCL is 5.
The signal to ADCH is further amplified by 16.5 times by the inverting amplifier circuit in Fig.~\ref{fig:analog_section}e and two times by a differential amplifier placed in before ADCH. 
The total gain of the circuit for ADCH is 165.
The ADC used for ADCL is LTC2325CUKG-12\#PBF which has $2~\mathrm{V_{pp}}$ and $12~\mathrm{bit}$ dynamic range.
The ADC used for ADCH is AD9637BCPZ-40 which has $2~\mathrm{V_{pp}}$ ane $12~\mathrm{bit}$ dynamic range. 

\begin{figure}[!t]
    \centering
    \includegraphics[width=0.8\columnwidth]{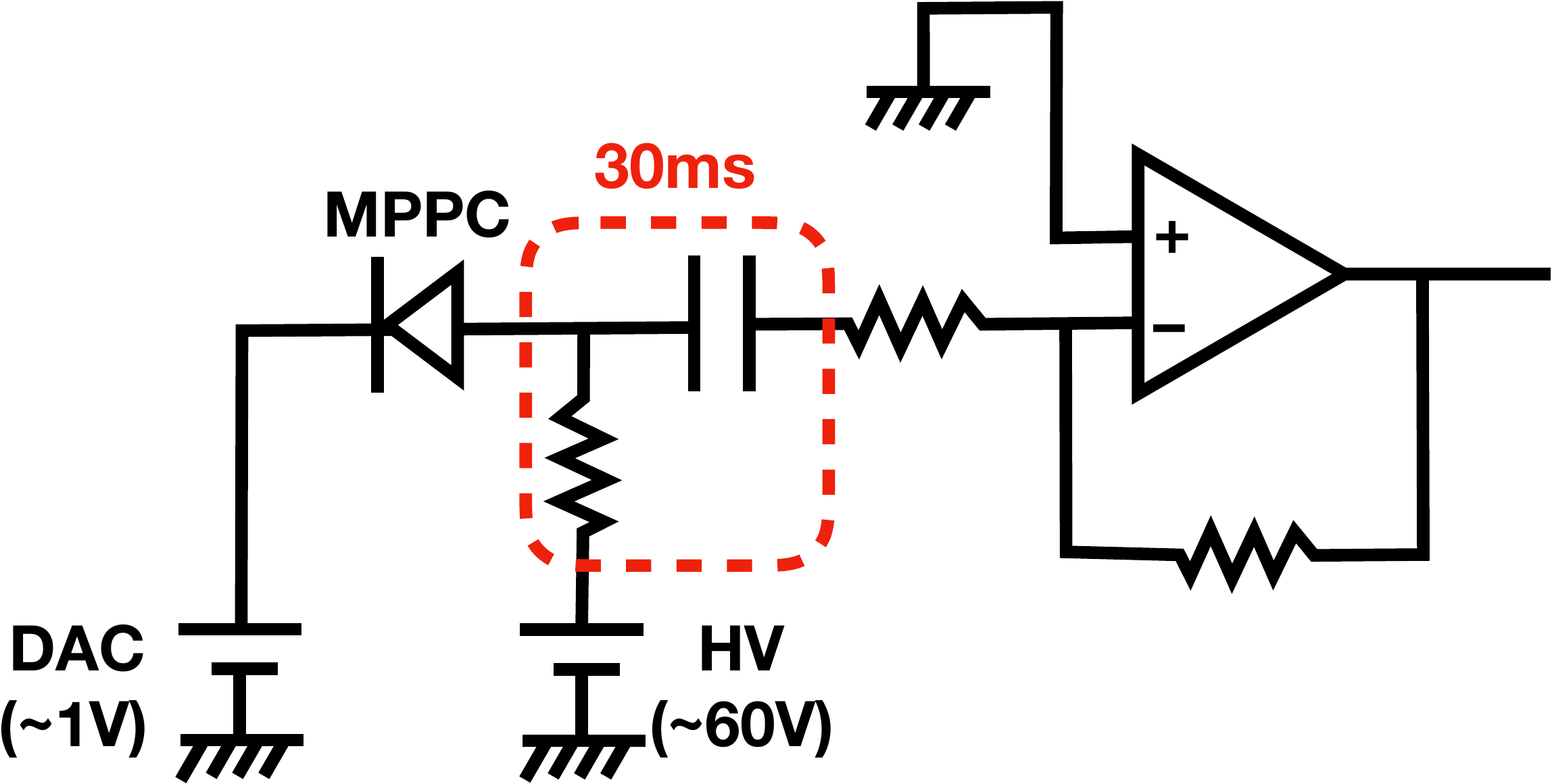}
    \caption{Diagram of AC coupling readout}
    \label{fig:ac_coupling}
\end{figure}

\begin{figure}[!t]
    \centering
    \includegraphics[width=\columnwidth]{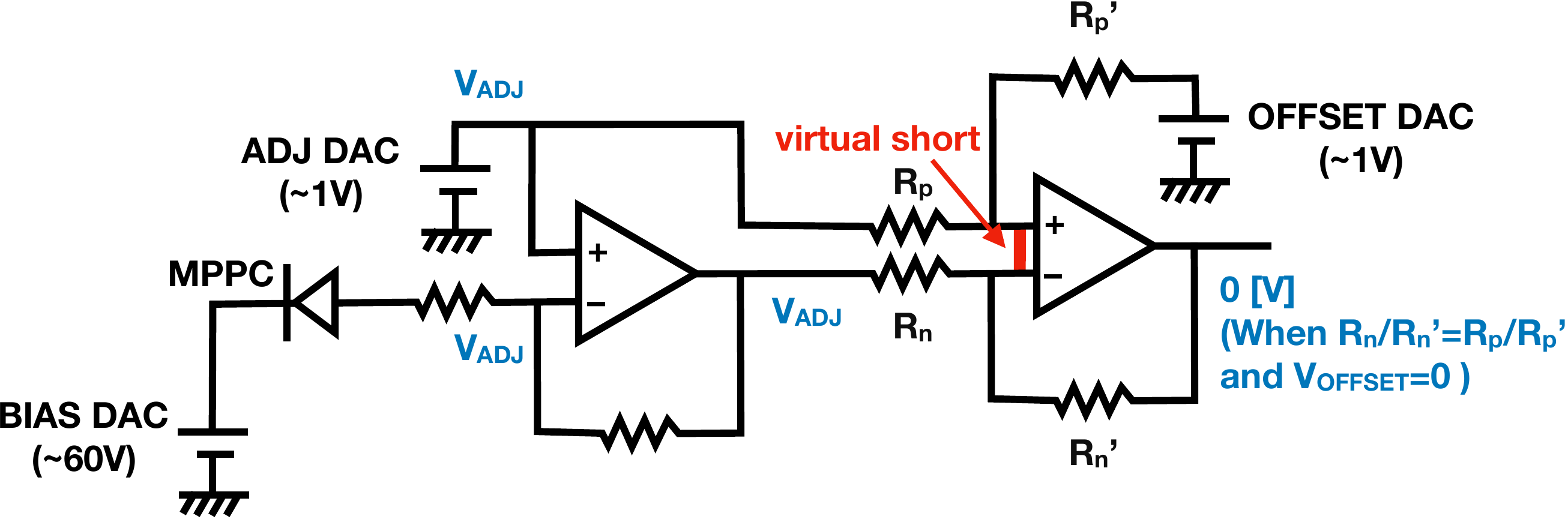}
    \caption{Diagram of the AxFEB DC coupling readout, which corresponds to the operational amplifiers of Fig.~\ref{fig:analog_section}c, first stage operational amplifiers of Fig.~\ref{fig:analog_section}d and Fig.~\ref{fig:analog_section}e).}
    \label{fig:dc_coupling}
\end{figure}

The bias voltage is applied by using two types of DAC's.
High voltage ($\sim 60~\mathrm{V}$) is generated by a DAC (LTC2630CSC6-LZ8) and applied to the cathode of all MPPC's, separated by low-pass filters.  
Then, a DAC (LTC2636CSC6-LZ8) is individually connected to the anode of each MPPC to finely adjust the bias voltage.
This scheme has been used for many readout board for SiPM. 
Since both electrodes are applied to finite voltage, the AC-coupling readout, as shown in Fig.~\ref{fig:ac_coupling} is often used. 
However, it is not adequate for our usage.
The maximum time width of $0\nu\beta\beta$ signal is $150\mathrm{~\mu s}$, and the required time constant of the low-pass filter is $30~\mathrm{ms}$ to suppress the deformation of the signal waveform.
Then, the acceptable event rate becomes very low.
To realize the DC-coupling readout, we adopted the configuration shown in Fig.~\ref{fig:dc_coupling}.
The bias voltage is adjusted by the $V_\mathrm{ADJ}$ via a virtual short of the operational amplifier.
Then, an offset corresponding to $V_\mathrm{ADJ}$ appears on the output of the operational amplifier.
This offset is canceled by connecting the $V_\mathrm{ADJ}$ to the second-stage amplifier via register $R_\mathrm{p}$.
When $R_\mathrm{n}/R_\mathrm{n}' = R_\mathrm{p}/R_\mathrm{p}'$ is satisfied, the offset of the output of the first stage operational amplifier is canceled.
An OFFSET DAC to the second operational amplifier controls the baseline dynamically.

\subsection{Digital section}
\label{fig:sec_digital_section}
The ADC chip for ADCL is  LTC2325CUKG-12\#PBF, whose dynamic range is 2~$\mathrm{V_{pp}}$ and sampling rate is $5 \mathrm{~MS/S}$.
By taking account of the gain of the analog section ($\times 5$), the maximum signal height of ADCL is about $400 \mathrm{~mV}$.
If the gain of MPPC is $2.6 \times 10^6$, the acceptable number of photons is $4,000 \mathrm{~photons}/200\mathrm{~\mu s}$.

The ADC chip for ADCH is AD9637BCPZ-40.
The dynamic range is $2~\mathrm{V_{pp}}$, and the sampling rate is $40\mathrm{~MS/s}$.
One ADCH exists for every seven channels, switched by a multiplexer.
The maximum signal height of ADCH is about 12~mV.

The AxFEB has a FPGA, Xilinx, and XC7A200T-1FBG484C.
It controls DACs (HV supply, bias voltage fine adjustment, offset adjustment) and multiplexer by SPI protocol and ADCs.
The ADC values are continuously written to the FIFO (ring buffer) created on the FPGA every clock.
When a trigger is issued, data writing is switched to another buffer, and the data in a previous buffer is sent to the PC using SiTCP~\cite{weko_188_1}.
The maximum data size for event for one board is about $2\mathrm{~Mbit}$.
To keep a certain margin, we chose the 200T series, which has $13\mathrm{~Mbit}$ memory.
The FPGA calculates the sum of 56~channels ADC counts, averages over three samples, and sends it to a general trigger module.
The trigger module calculates the total sum of the data sent from AxFEB's and issues a trigger based on the sum value.

AxFEB can be driven by either internal or external clocks.
When using several boards at the same time, clocks are supplied from the trigger module.
The communication between AxFEB and the trigger module is conducted by LVDS ($160\mathrm{~MHz}$).
The ADC sum of 56~channels is converted to a pseudo analog signal by DAC ($1\mathrm{~MS/s}$), the output from a LEMO connector, and can be checked by an oscilloscope.

Data transportation and slow control between AxFEB and PC are performed via Ethernet.
Acquired data is transported by TCP protocol, and AxFEB is slow controlled by UDP protocol.
In order to perform this communication, we use SiTCP as a technology to connect FPGA and Ethernet.
The IP address of AxFEB can be set with an $8\mathrm{~bit}$ DIP switch on the board.

\section{Performance Evaluation}
\begin{figure}[!t]
    \centering
    \includegraphics[width=0.8\linewidth]{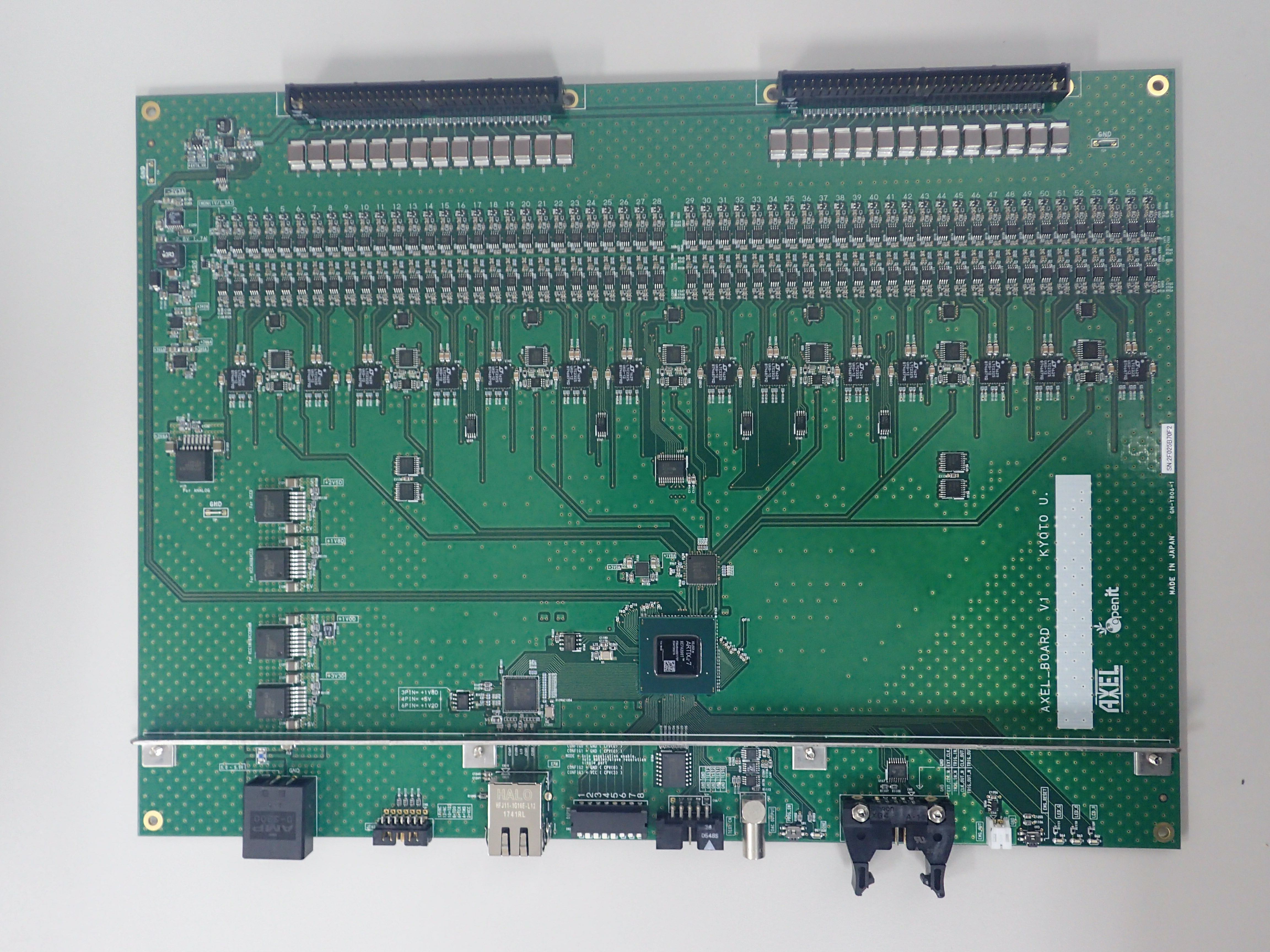}
    \caption{Photograph of AxFEB}
    \label{fig:AxFEB}
\end{figure}

Fig.~\ref{fig:AxFEB} shows a picture of AxFEB.

\subsection{Performances of ADCL}
\begin{figure}[!t]
    \centering
    \includegraphics[width=\linewidth]{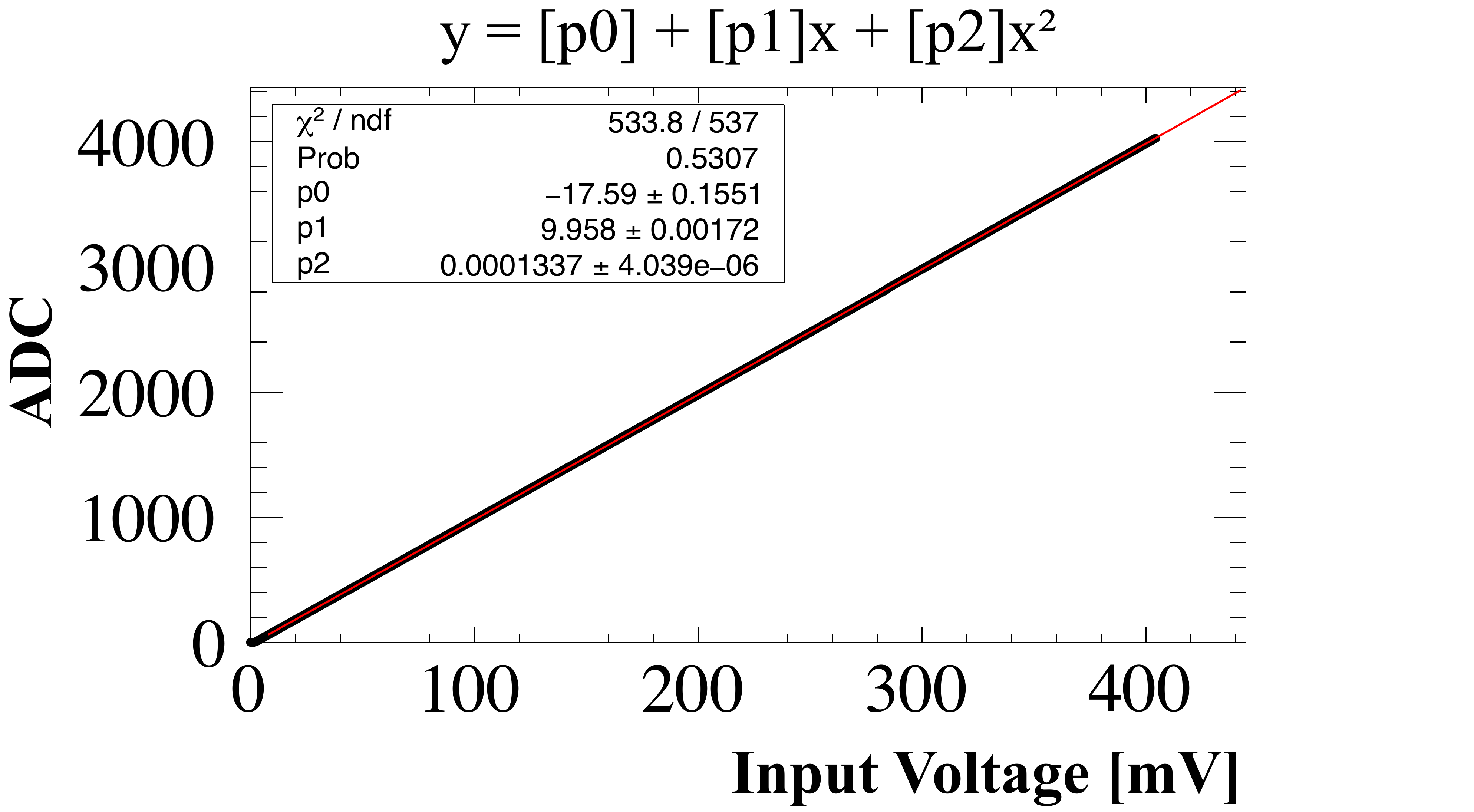}
    \caption{Example of the result of ADCL linearity test}
    \label{fig:ADCL_linearity}
\end{figure}

The linearity of ADCL was checked by inputting constant voltage to AxFEB with $1\mathrm{~mV}$ step.
Fig.~\ref{fig:ADCL_linearity} shows the measured ADC value with respect to the input voltage for a certain channel.
It was fitted with a quadratic function. 
The second-order term is about $10^{-5}$ of the first-order term.
The non-linearity of ADCL is sufficiently small.
We also calculate the gain of the analog section of ADCL for 56~channels.
The gain of the analog section was also measured and confirmed that the value is as designed, $\times \sim 5$.

\begin{figure}[!t]
    \centering
    \includegraphics[width=1.0\linewidth]{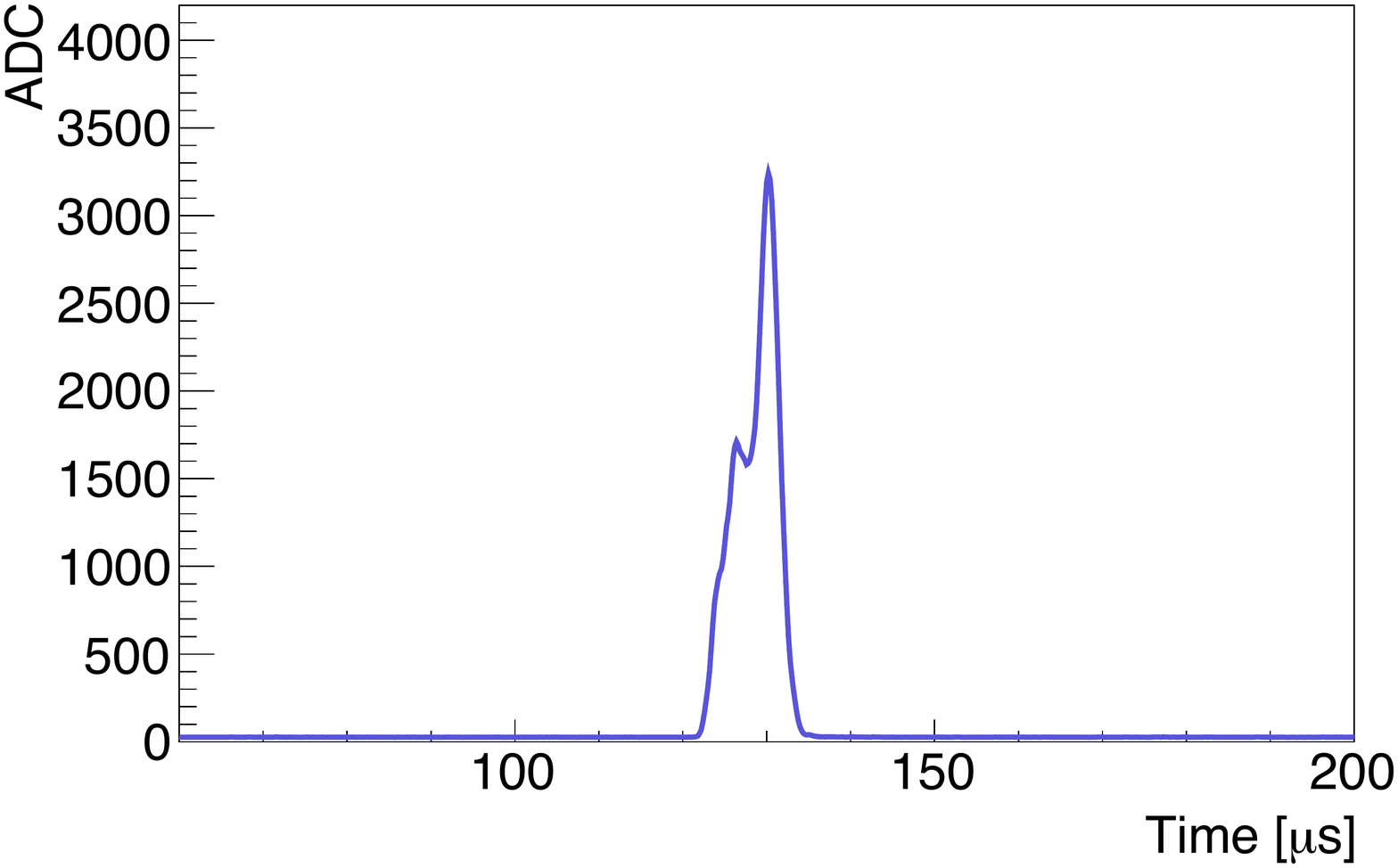}
    \caption{Example of the waveform of one channel obtained by AxFEB. The 180-L size prototype detector is irradiated with a $^{22}$Na source.}
    \label{fig:el_waveform}
\end{figure}

Fig.~\ref{fig:el_waveform} shows an example of the waveform of one channel obtained by AxFEB when the 180-L size prototype detector is irradiated with a $^{22}$Na source.

\subsection{Performance of ADCH}
\begin{figure}[!t]
    \centering
    \includegraphics[width=\linewidth]{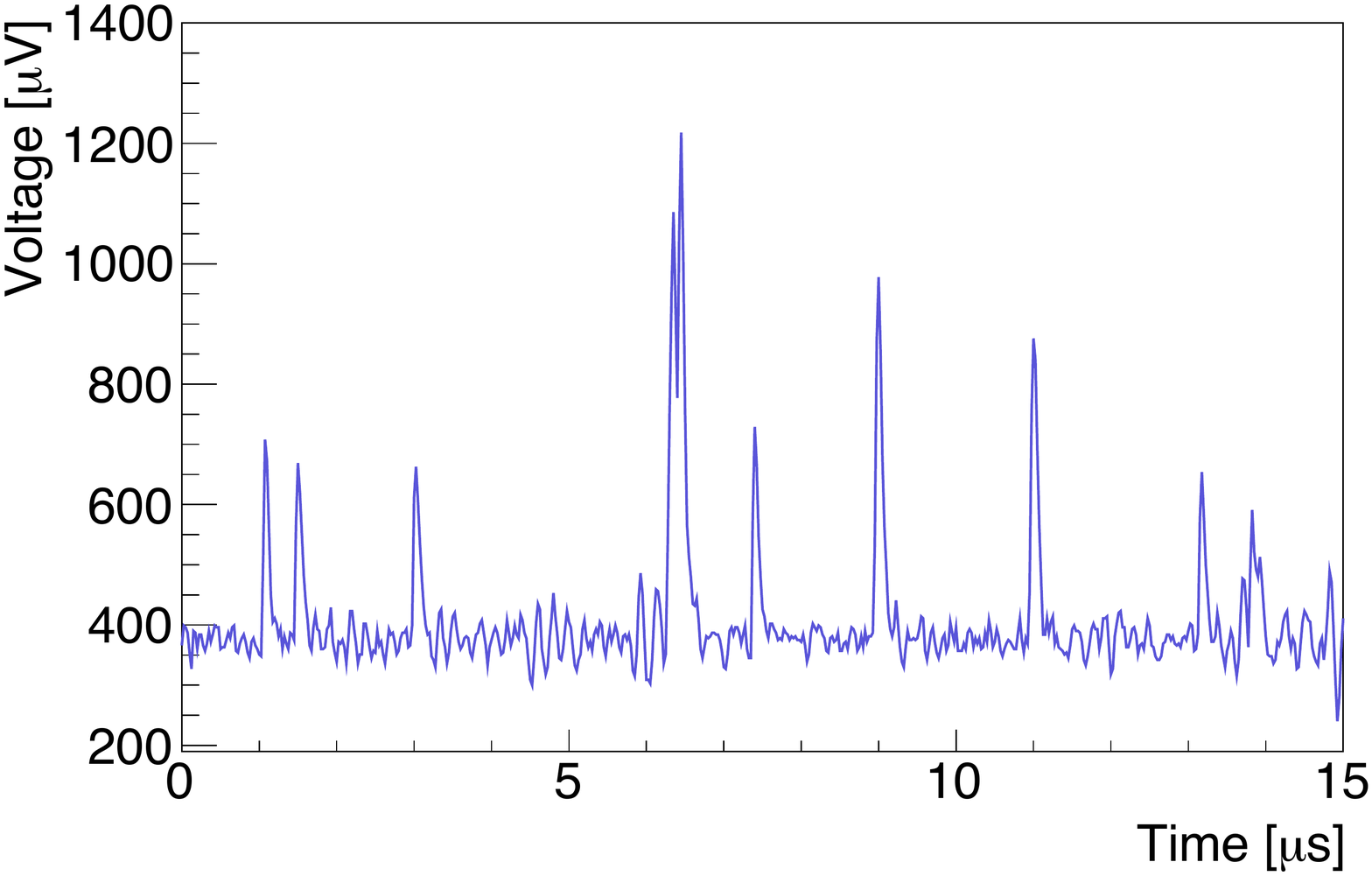}
    \caption{Example waveform taken by ADCH. Dark current pulses corresponding to $1~\mathrm{p.e.}$ and $2~\mathrm{p.e.}$ are clearly seen.}
    \label{fig:ADCH_waveform}
\end{figure}

\begin{figure}
    \centering
    \includegraphics[width=1.0\linewidth]{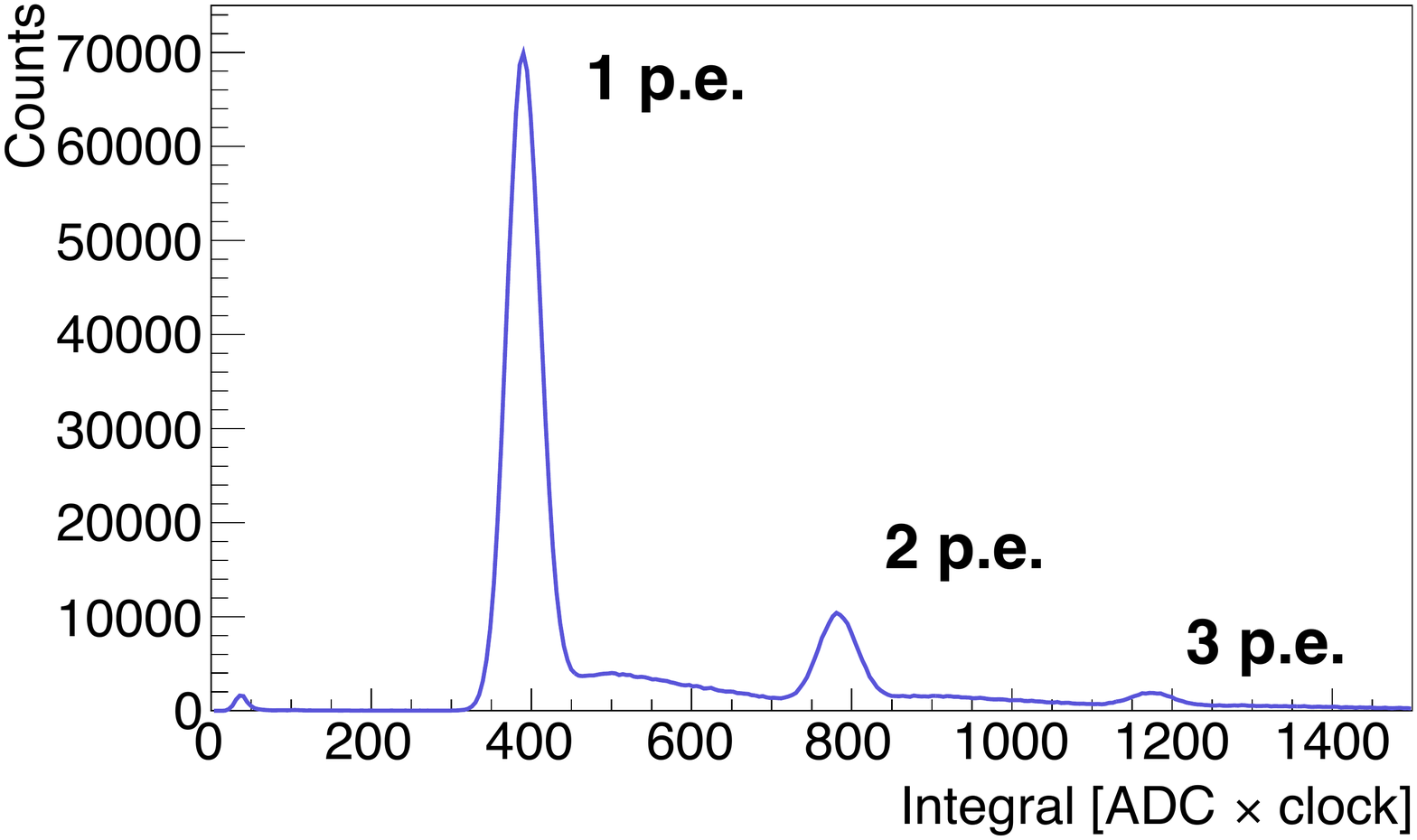}
    \caption{Dark current pulse integral distribution obtained by ADCH.
    Peaks corresponding to $1~\mathrm{p.e.}$, $2~\mathrm{p.e.}$, and $3~\mathrm{p.e.}$ are clearly separated.
    Those between peaks are caused by after pulses of MPPC.}
    \label{fig:dark_distribution_ADCH}
\end{figure}

Fig.~\ref{fig:ADCH_waveform} shows an example waveform acquired by ADCH.
The pulses were integrated and the distribution were obtained as shown in Fig.~\ref{fig:dark_distribution_ADCH}.
Dark current pulses corresponding to $1~\mathrm{p.e.}$ and $2~\mathrm{p.e.}$ are clearly separated.

\begin{figure}[!t]
    \centering
    \includegraphics[width=\linewidth]{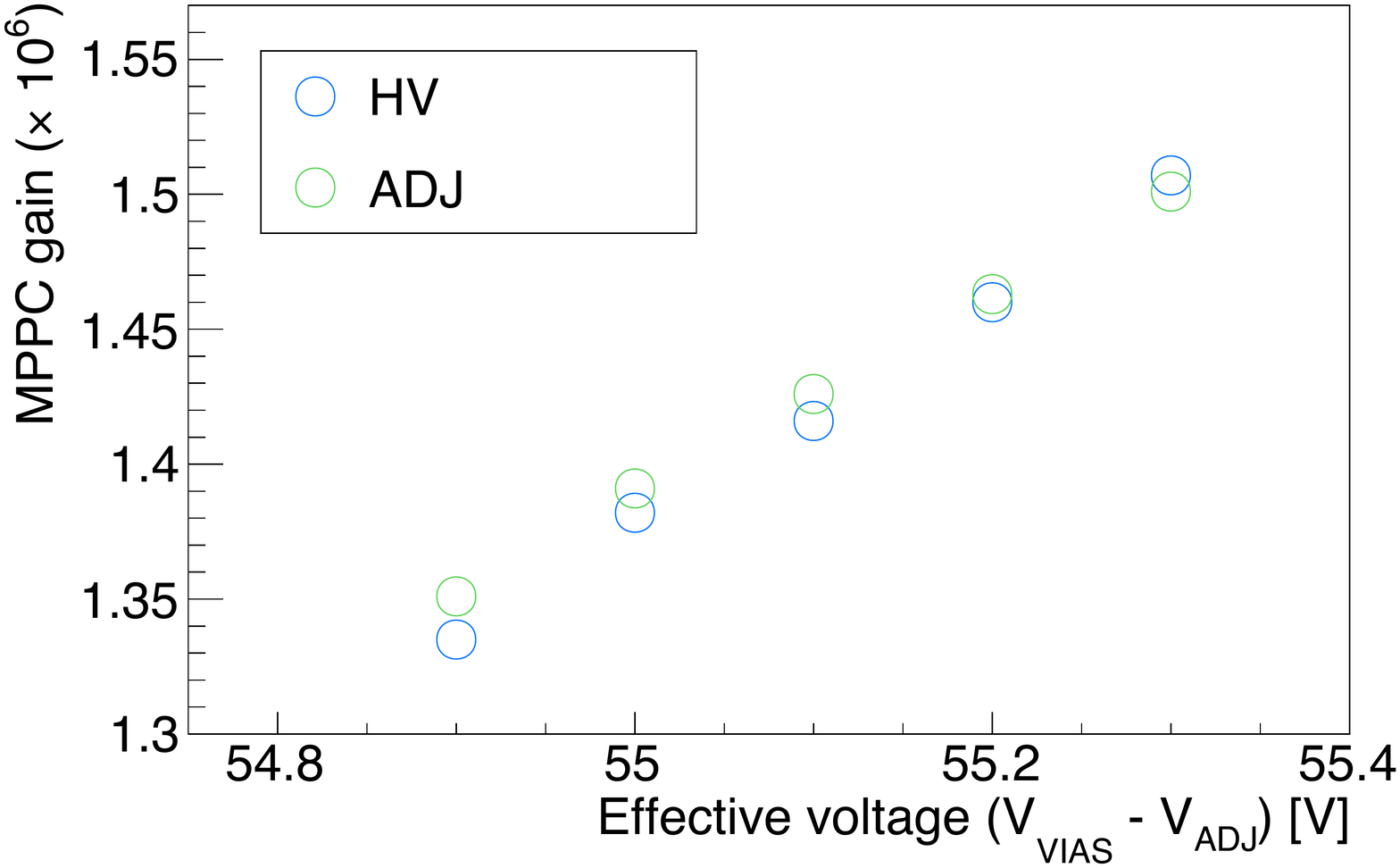}
    \caption{Measured MPPC gain as a function of the bias voltage. The blue points represent the gain when the common bias voltage (BIAS) is changed from $54.9\mathrm{~V}$ to $55.3\mathrm{~V}$, while $\mathrm{V_{ADJ}}$ is fixed to zero. The green points represent the gain when the $\mathrm{V_{ADJ}}$ is changed with fixing the the common bias voltage to $55.3\mathrm{~V}$.}
    \label{fig:gain_adjustment}
\end{figure}

Fig.~\ref{fig:gain_adjustment} shows the measured MPPC gain as a function of the bias voltage.
The blue points represent the gain when the common bias voltage (BIAS) is changed from $54.9\mathrm{~V}$ to $55.3\mathrm{~V}$, while $\mathrm{V_{ADJ}}$ is fixed to zero. The green points represent the gain when the $\mathrm{V_{ADJ}}$ is changed with fixing the common bias voltage to $55.3\mathrm{~V}$.
From the result, it was confirmed that the gain can be controlled by the applied voltage expressed by $\mathrm{V_{BIAS}}-\mathrm{V_{ADJ}}$.

The gain of MPPC changes by about $2.5\%$ with $0.1\mathrm{~V}$ bias voltage change.
Since the voltage of the DAC can be adjusted in $0.01\mathrm{~V}$ steps, the gain of each MPPC can be adjusted by about $0.25\%$.

\begin{figure}[!t]
    \centering
    \includegraphics[width=\linewidth]{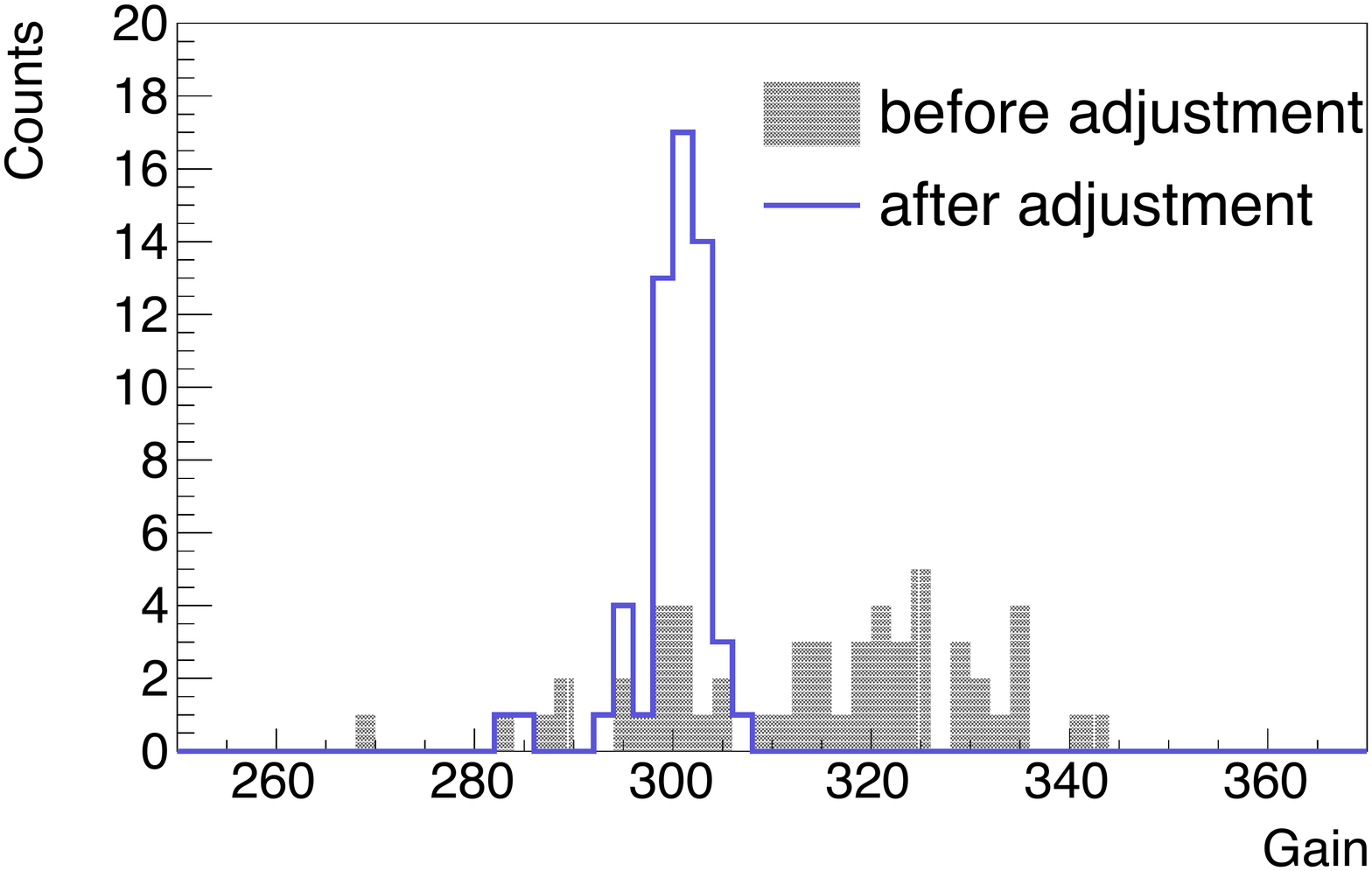}
    \caption{MPPC gain distribution before/after the bias voltage adjustment}
    \label{fig:gain_before_after}
\end{figure}

Fig.~\ref{fig:gain_before_after} shows a result of the fine adjustment of the bias voltage for 56-channels of MPPC's.
The gains of individual MPPC's, were different when only common bias voltage was applied; on the other hand, gains are aligned after fine adjustment was conducted.

\section{Summary}
We are developing a high-pressure xenon gaseous TPC for searching for neutrinoless double-beta decay.
The ionization signal is readout by detecting electroluminescence photons with SiPM's.  
A large number of channels have to be readout at $5~\mathrm{MS/s}$ with a wide dynamic range. 
Currently, we are constructing a 180-L size prototype detector to demonstrate the detector performance at the Q-value and to establish the techniques for larger-size detector.
The number of SiPM's for this detector is 1,512~channels.
To operate and readout such many SiPM's, we have developed a front-end electronics board as AxFEB.
The bias voltage to SiPM's can be adjusted for individual channels.
The signal is readout with the DC coupling by canceling the offset caused by bias adjustment. 
To calibrate and monitor the gain of SiPM's, the board has an additional high gain ADC to measure the dark current.
We evaluated the performance of AxFEB and confirmed that it satisfies the requirement.

\section*{Acknowledgment}
We thank the Open-It (Open source consortium for Instrumentation) for grateful support and advice in developing electronics.
We also express our thanks to other members of the AXEL collaboration.

\bibliographystyle{unsrt}

\end{document}